\documentclass{elsart}
\usepackage{epsfig}
\usepackage{amssymb}
\usepackage{amsmath}

\newcommand{\affuni}[2]{Dipartimento di Fisica dell'Universit\`a #1, #2, Italy.}
\newcommand{\affinfn}[2]{INFN Sezione di #1, #2, Italy.}

\newcommand{\dafne}     {DA$\Phi$NE }

\newcommand{\ep}{\mbox{$e^{+}$}}
\newcommand{\el}{\mbox{$e^{-}$}}
\newcommand{\pio}{\mbox{$\pi^{0}$}}

\newcommand{\ks}{\mbox{$K_{S}$}}

\begin{document}
%%%%%%%%%%%%%%%%%%%%%%%%%%%%%%%%%%%%%%%%%%%%%%%%%%%%%%%%%%%%%%%%%%%%%%%%%%%%%%%

%==============================================================================
\begin{frontmatter}
%==============================================================================

\title{\boldmath Study of Dalitz decay $\phi \to \eta  e^+e^-$ with KLOE detector}

\collab{The KLOE-2 Collaboration}
%\author[Roma2,INFNRoma2]{F.~Archilli},
\author[Frascati]{D.~Babusci},
%\author[Roma2,INFNRoma2]{D.~Badoni},
\author[Cracow]{I.~Balwierz-Pytko},
\author[Frascati]{G.~Bencivenni},
%\author[Roma1,INFNRoma1]{C.~Bini},
\author[Frascati]{C.~Bloise},
%\author[INFNRoma1]{V.~Bocci},
\author[Frascati]{F.~Bossi},
\author[INFNRoma3]{P.~Branchini},
\author[Roma3,INFNRoma3]{A.~Budano},
%\author[Moscow]{S.~A.~Bulychjev},
\author[Uppsala]{L.~Caldeira~Balkest\aa hl},
%\author[Frascati]{P.~Campana},
%\author[Frascati]{G.~Capon},
\author[Roma3,INFNRoma3]{F.~Ceradini},
\author[Frascati]{P.~Ciambrone},
\author[Messina,INFNCatania]{F.~Curciarello},
\author[Cracow]{E.~Czerwi\'nski},
\author[Frascati]{E.~Dan\`e},
\author[Messina,INFNCatania]{V.~De~Leo},
\author[Frascati]{E.~De~Lucia},
\author[INFNBari]{G.~De~Robertis},
%\author[Roma1,INFNRoma1]{A.~De~Santis},
\author[Frascati]{A.~De~Santis},
%\author[Roma1,INFNRoma1]{G.~De~Zorzi},
\author[Frascati]{P.~De~Simone},
\author[Roma3,INFNRoma3]{A.~Di~Cicco},
\author[Roma1,INFNRoma1]{A.~Di~Domenico},
%\author[Napoli,INFNNapoli]{C.~Di~Donato},
\author[INFNRoma2]{R.~Di~Salvo},
%\author[Roma3,INFNRoma3]{B.~Di~Micco},
\author[Frascati]{D.~Domenici},
%\author[Frascati]{A.~D'Uffizi},
\author[Bari,INFNBari]{O.~Erriquez},
\author[Bari,INFNBari]{G.~Fanizzi},
\author[Roma2,INFNRoma2]{A.~Fantini},
\author[Frascati]{G.~Felici},
\author[ENEACasaccia,INFNRoma1]{S.~Fiore},
\author[Roma1,INFNRoma1]{P.~Franzini},
\author[Cracow]{A.~Gajos},
\author[Roma1,INFNRoma1]{P.~Gauzzi},
\author[Messina,INFNCatania]{G.~Giardina},
\author[Frascati]{S.~Giovannella\corauthref{cor}},
\ead{simona.giovannella@lnf.infn.it}
%\author[Roma2,INFNRoma2]{F.~Gonnella},
\author[INFNRoma3]{E.~Graziani},
\author[Frascati]{F.~Happacher},
\author[Uppsala]{L.~Heijkenskj\"old}
\author[Uppsala]{B.~H\"oistad},
%\author[Frascati]{L.~Iafolla},
%\author[Energetica,Frascati]{E.~Iarocci},
%\author[Uppsala]{M.~Jacewicz},
\author[Uppsala]{T.~Johansson},
%\author[Cracow]{K.~Kacprzak},
\author[Cracow]{D.~Kami\'nska},
\author[Cracow]{W.~Krzemien},
%\author[Warsaw]{A.~Kowalewska},
%\author[Moscow]{V.~Kulikov},
\author[Uppsala]{A.~Kupsc},
\author[Frascati,StonyBrook]{J.~Lee-Franzini},
%\author[Frascati]{B.~Leverington},
\author[INFNBari]{F.~Loddo},
\author[Roma3,INFNRoma3]{S.~Loffredo},
\author[Messina,INFNCatania,CentroCatania]{G.~Mandaglio},
\author[Moscow]{M.~Martemianov},
\author[Frascati,Marconi]{M.~Martini},
\author[Roma2,INFNRoma2]{M.~Mascolo},
%\author[Moscow]{M.~Matsyuk},
\author[Roma2,INFNRoma2]{R.~Messi},
\author[Frascati]{S.~Miscetti\corauthref{cor}},
\ead{stefano.miscetti@lnf.infn.it}
\author[Frascati]{G.~Morello},
\author[INFNRoma2]{D.~Moricciani},
\author[Cracow]{P.~Moskal},
%\author[INFNRoma3,LIP]{F.~Nguyen},
%\author[Frascati]{L.~Quintieri},
\author[Frascati]{A.~Palladino},
\author[INFNRoma3]{A.~Passeri},
\author[Energetica,Frascati]{V.~Patera},
\author[Roma3,INFNRoma3]{I.~Prado~Longhi},
\author[INFNBari]{A.~Ranieri},
%\author[Mainz]{C.~F.~Redmer},
\author[Frascati]{P.~Santangelo},
\author[Frascati]{I.~Sarra\corauthref{cor}},
\ead{ivano.sarra@lnf.infn.it}
\corauth[cor]{Corresponding author.}
\author[Calabria,INFNCalabria]{M.~Schioppa},
\author[Frascati]{B.~Sciascia},
%\author[Energetica,Frascati]{A.~Sciubba},
\author[Cracow]{M.~Silarski},
%\author[Calabria,INFNCalabria]{S.~Stucci},
%\author[Roma3,INFNRoma3]{C.~Taccini},
\author[INFNRoma3]{L.~Tortora},
\author[Frascati]{G.~Venanzoni},
%\author[Frascati,CERN]{R.~Versaci},
\author[Warsaw]{W.~Wi\'slicki},
\author[Uppsala]{M.~Wolke}
%\author[Cracow]{J.~Zdebik}
%%%%%%%%%%%%%%%%%%%%%%%%%%%%%%%%%%%%%%%%%%%%%%%%%%%%%%%%%%%%%%%%%%%%%%%%%%%%%%%%%%%%%%%%%%%%%%%%%%
\address[Bari]{\affuni{di Bari}{Bari}}
\address[INFNBari]{\affinfn{Bari}{Bari}}
\address[CentroCatania]{Centro Siciliano di Fisica Nucleare e Struttura della Materia, Catania, Italy.}
\address[INFNCatania]{\affinfn{Catania}{Catania}}
\address[Calabria]{\affuni{della Calabria}{Cosenza}}
\address[INFNCalabria]{INFN Gruppo collegato di Cosenza, Cosenza, Italy.}
\address[Cracow]{Institute of Physics, Jagiellonian University, Cracow, Poland.}
\address[Frascati]{Laboratori Nazionali di Frascati dell'INFN, Frascati, Italy.}
%\address[Messina]{\affuni{di Messina}{Messina}}
%\address[Mainz]{Institut f\"ur Kernphysik, 
%Johannes Gutenberg Universit\"at Mainz, Germany.}
\address[Messina]{Dipartimento di Fisica e Scienze della Terra dell'Universit\`a di Messina, Messina, Italy.}\address[Moscow]{Institute for Theoretical and Experimental Physics (ITEP), Moscow, Russia.}
%\address[Napoli]{\affuni{``Federico II''}{Napoli}}
%\address[INFNNapoli]{\affinfn{Napoli}{Napoli}}
\address[Energetica]{Dipartimento di Scienze di Base ed Applicate per l'Ingegneria dell'Universit\`a 
``La Sapienza'', Roma, Italy.}
\address[Marconi]{Dipartimento di Scienze e Tecnologie applicate, Universit\`a ``Guglielmo Marconi", Roma, Italy.}
\address[Roma1]{\affuni{``La Sapienza''}{Roma}}
\address[INFNRoma1]{\affinfn{Roma}{Roma}}
\address[Roma2]{\affuni{``Tor Vergata''}{Roma}}
\address[INFNRoma2]{\affinfn{Roma Tor Vergata}{Roma}}
\address[Roma3]{Dipartimento di Matematica e Fisica dell'Universit\`a 
``Roma Tre'', Roma, Italy.}
%\address[Roma3]{\affuni{``Roma Tre''}{Roma}}
\address[INFNRoma3]{\affinfn{Roma Tre}{Roma}}
\address[ENEACasaccia]{ENEA UTTMAT-IRR, Casaccia R.C., Roma, Italy}
\address[StonyBrook]{Physics Department, State University of New 
York at Stony Brook, USA.}
\address[Uppsala]{Department of Physics and Astronomy, Uppsala University, Uppsala, Sweden.}
\address[Warsaw]{National Centre for Nuclear Research, Warsaw, Poland.}

%------------------------------------------------------------------------------

\begin{abstract}

%The conversion decays  $\rm{\phi \to \eta e^+e^-}$, with $\rm{\eta\to\pi^0\pi^0\pi^0}$, was studied at KLOE detector using $\phi$-meson production in $\rm{e^+e^-}$ annihilation %at DA$\Phi$NE collider. The branching ratios of this decays was measured: $\rm{BR(\phi \to \eta e^+e^-) = }$$\rm{(1.075 \pm 0.007 \pm 0.038 ^{+0.006}_{-0.002})}$$\rm{ \times %10^{-4}}$. The $\rm{e^+e^-}$ pair mass spectra and transition form factors were also studied, extracting
%the slope of the transition form factor: $\rm{b_{\phi \eta} = }$$\rm{(1.17 \pm 0.10 ^{+0.07}_{-0.11})\,GeV^{-2}}$
We have studied the vector to pseudoscalar conversion decay \textit{$\phi \to \eta e^+e^-$}, with $\rm{\eta\to\pi^0\pi^0\pi^0}$, with the KLOE detector at DA$\Phi$NE. The data set of 1.7 fb$^{-1}$ of $e^+e^-$ collisions at $\rm{\sqrt s \sim M_{\phi}}$ contains a clear conversion decay signal of $\rm{\sim 31,000}$ events from which we measured a value of $\rm{BR(\phi \to \eta \mathit{e^+e^-}}) = $$\rm{(1.075 \pm 0.007 \pm 0.038)}$$\rm{ \times 10^{-4}}$. 
%Motivated by puzzling deviation from vector meson dominance theory in the $\omega \to \pi^0 e^+e^-$, 
The same sample is used to determine the transition form factor by a fit to the $e^+e^-$ invariant mass spectrum, obtaining $\rm{b_{\phi \eta} = }$$\rm{(1.17 \pm 0.10 ^{+0.07}_{-0.11})\,GeV^{-2}}$, that improves by a factor of five the precision of the previous measurement and is in good agreement with VMD expectations.
%and the derived transition form factor $|F|^2$. We extract a value for the slope of $\rm{b_{\phi \eta} = }$$\rm{(1.17 \pm 0.10 ^{+0.07}_{-0.11})\,GeV^{-2}}$, that improves by a factor of five the precision of the previous measurement and is in good agreement with VMD expectations.

\end{abstract}

%------------------------------------------------------------------------------


\end{frontmatter}

%==============================================================================
\section{Introduction}
%==============================================================================
\label{Sec:Intro}

We report the study of the vector to pseudoscalar conversion decay $\phi \to \eta e^+e^-$ with $\eta\to\pi^0\pi^0\pi^0$. In conversion decays, $A \to B\gamma^*
\to B\,e^+e^-$, the radiated photon is virtual and the squared dilepton invariant mass, $M_{ee}^2$, corresponds to the photon 4-momentum transferred, $q^2$. The probability of having a lepton pair of given invariant mass is determined by the electromagnetic dynamical structure of the transition $A \to B \gamma^*$. 
The differential decay rate, normalized to the radiative width, is ~\cite{biblio:Landsberg854}:
%The nature of the mesons and their underlying quark and gluon structure can be described by the transition form factor $\rm{F_{AB}(q^2)}$, obtained from the probability of the conversion decay as a function of the invariant mass of the lepton pair. It can be parametrizated, experimentally, as~\cite{biblio:Landsberg854}:
\begin{eqnarray}
\label{eq:fit_func}
  %\frac{d}{dq^2}\frac{\Gamma(\phi\to\eta\,e^+e^-)}{\Gamma(\phi\to\eta\gamma)}
  \frac{1}{\Gamma(\phi\to\eta\gamma)}\frac{d\Gamma(\phi\to\eta\,e^+e^-)}{dq^2}
  = \ \ \ \ \ \ \ \ \ \ \ \ \ \ \ \ \ \ \ \ \ \ 
  \ \ \ \ \ \ \ \ \ \ \ \ \ \ \ \ \ \ \ \ \ \
  \ \ \ \ \ \ \ \ \ \ \ \ \ \ \ \ \ \ \ \ \ \ 
  \ \ \ \ \ \ \ \ \ \ \ \nonumber \\
  \frac{\alpha}{3\pi}\frac{|F_{\phi\eta}(q^2)|^2}{q^2}
  \sqrt{1-\frac{4M^2}{q^2}} \left(1+\frac{2M^2}{q^2}\right)
  \left[\left(1+\frac{q^2}{M_\phi^2-M_\eta^2}\right)^2 -
  \frac{4 M_\phi^2 q^2}{(M_\phi^2-M_\eta^2)^2} \right]^{3/2},\,\ \ \ \ \ \ \ \ \ \ \ \ \ \ \nonumber
\end{eqnarray}
where $m$ is the mass of the electron and $M_{\phi}$, $M_{\eta}$ are the masses of the $\phi$ and $\eta$ mesons, respectively. $F_{\phi\eta}(q^2)$ is the transition form factor, TFF,  that describes the coupling of the mesons to virtual photons and provides information on its nature and underlying structure. The slope of the transition form factor, $b_{\phi\eta}$, is defined as:
\begin{equation}
\label{eq:lambda}
 b_{\phi \eta} \equiv \frac{dF}{dq^2}|_{q^2=0}.
\end{equation}
In the Vector Meson Dominance model, VMD, the transition form factor is parametrized as:
\begin{equation}
\label{eq:lambda_vmd}
  F_{\phi\eta}(q^2) = \frac{1}{1-q^2/\Lambda_{\phi\eta}^2} \;\;\rightarrow\;\;  b_{\phi \eta} \approx \Lambda_{\phi\eta}^{-2}.
  \end{equation}
The VMD successfully describes some transitions, such as  $\eta \to \gamma \mu^+\mu^-$, while is failing for others, as in the case of $\omega \to \pi^0 \mu^+\mu^-$ \cite{biblio:eta_mumu}. Recently, new models have been developed to overcome such a kind of discrepancies~\cite{biblio:Leupold,biblio:Ivashyn} and they should be validated with the experimental data from other channels. The only existing data on $\phi \to \eta e^+e^-$ come from the SND~\cite{biblio:snd4} and CMD-2~\cite{biblio:cmd4} experiments. Their measurements of the branching ratio, BR$(\,\phi\to\eta e^+e^-)$, are (1.19$\pm$0.19$\pm$0.07)$\times 10^{-4}$ and (1.14$\pm$0.10$\pm$0.06)$\times 10^{-4}$, respectively. The VMD expectation is  BR$(\,\phi\to\eta e^+e^-\,)$ = 1.1$\times 10^{-4}$ \cite{biblio:vmd_expectation}. The SND experiment has also measured the slope of the transition form factor from the $M_{ee}$ invariant mass distribution, on the basis of 213 events: $b_{\phi \eta}$= (3.8$\pm$1.8) GeV$^{-2}$ \cite{biblio:snd4}. The VMD expectation is $b_{\phi \eta}$=1 GeV$^{-2}$ \cite{biblio:vmd_expectation}.\\
%The value of the transition form factor fixes also the upper limit for the light dark boson searches in $\phi \to \eta e^+e^-$.
%It confirms the exclusion plot for the U boson in the mass range 5 $<$ M$_U$ $<$ 470 MeV, from 1.7 $\times$ 10$^{-5}$ and 8.0 $\times$ 10$^{-6}$, obtained by combining both samples ($\rm{\eta \to \pi^+\pi^-\pi^0}$ and $\rm{\eta \to \pi^0\pi^0\pi^0}$) assuming a $b_{\phi\eta}$ = 1 GeV$^{-2}$.
Due to the large data sample, we have performed three different measurements:
\begin{enumerate} 
\item the determination of the branching fraction of the $\phi \to \eta e^+e^-$ decay;
\item the direct measurement of the transition form factor slope b$_{\phi \eta}$ with a fit to the dilepton invariant mass spectrum;
\item the extraction of the $|F_{\phi\eta}|^2$ as a function of the dilepton invariant mass.
\end{enumerate}

%The only existing data on $\rm{\phi \to \eta e^+e^-}$ come from the SND~\cite{biblio:snd4} and CMD-2~\cite{biblio:cmd4} experiments. Their BR$(\rm{\,\phi\to\eta e^+e^-\,)}$ measurements are $\rm{(1.19 \pm 0.19 \pm 0.07)\times 10^{-4}}$ and $\rm{(1.14 \pm 0.10 \pm 0.06)\times 10^{-4}}$ respectively. \\
%The VMD expectation is  BR$(\rm{\phi\to\eta e^+e^-\,)}$ = 1.1$\times 10^{-4}$ \cite{biblio:vmd_expectation}. \\
%SND experiment has also measured the transition form factor from the $M_{ee}$ invariant mass distribution on the basis of 213 events: $\rm{b_{\phi \eta}}$=3.8$\pm$1.8 GeV$^{-2}$ \cite{biblio:snd4}. The VMD expectation is $\rm{b_{\phi \eta}}$=1 GeV$^{-2}$ \cite{biblio:vmd_expectation}.

%==============================================================================
\section{The KLOE detector}
%==============================================================================

\dafne, the Frascati $\phi$-factory, is an $e^+e^-$ collider running at 
center of mass energy of $\sim 1020$~MeV. %, the mass of the $\phi$ meson. 
Positron and electron beams collide at an angle of $\pi$-25 mrad, 
producing $\phi$ mesons nearly at rest. The KLOE experiment operated at 
this collider from 2000 to 2006, collecting 2.5 fb$^{-1}$.
The KLOE apparatus consists of a large cylindrical Drift Chamber 
surrounded by a lead-scintillating fiber electromagnetic calorimeter 
both inserted inside a superconducting coil, providing a 0.52~T 
axial field.
The beam pipe at the interaction region is a sphere with 10 cm radius, 
made of a 0.5 mm thick Beryllium-Aluminum alloy. 
The drift chamber~\cite{biblio:DCH}, 4~m in diameter and 3.3~m long, has 
12,582 all-stereo tungsten sense wires and 37,746 aluminum field wires,
with a shell made of carbon fiber-epoxy composite with an internal wall 
of $\sim 1$ mm thickness. The gas used is a 90\% helium, 10\% isobutane 
mixture. The momentum resolution is $\sigma(p_{\perp})/p_{\perp}\approx 0.4\%$.
Vertices are reconstructed with a spatial resolution of $\sim$ 3~mm.
The calorimeter~\cite{biblio:EMC}, with a readout granularity of 
$\sim$\,(4.4 $\times$ 4.4)~cm$^2$, for a total of 2440 cells arranged 
in five layers, covers 98\% of the solid angle. Each cell is read out 
at both ends by photomultipliers, both in amplitude and time. 
The energy deposits are obtained from the signal amplitude while the
arrival times and the particles positions are obtained from the time
differences. 
Cells close in time and space are grouped into energy clusters. 
%The cluster energy $E$ is the sum of the cell energies.
%The cluster time $T$ and position $\vec{R}$ are energy-weighted averages. 
Energy and time resolutions are $\sigma_E/E = 5.7\%/\sqrt{E\ {\rm(GeV)}}$ 
and  $\sigma_t = 57\ {\rm ps}/\sqrt{E\ {\rm(GeV)}} \oplus100\ {\rm ps}$, 
respectively.
The trigger \cite{biblio:TRG} uses both calorimeter and chamber information.
In this analysis the events are selected by the calorimeter trigger,
requiring two energy deposits with $E>50$ MeV for the barrel and $E>150$
MeV for the endcaps.\\ 
Machine parameters are measured online by means of large angle Bhabha scattering events. 
The average value of the center of mass 
energy is evaluated with a precision of about 30 keV each 200 nb$^{-1}$ of integrated luminosity.
Collected data are processed by an event classification algorithm~\cite{biblio:NIMOffline},
which streams various categories of events in different output files.

%==============================================================================
\section{\boldmath Branching Ratio}
%==============================================================================
\label{Sec:DataSample}

\begin{figure}[!t]
  \begin{center}
    \epsfig{file=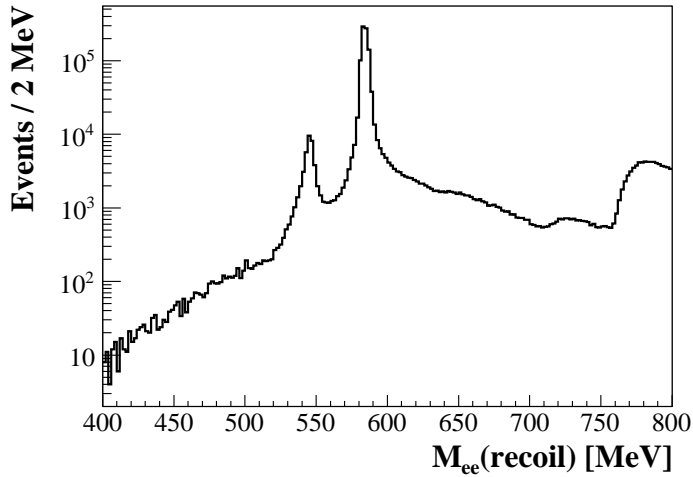,width=0.8\textwidth}
  \end{center}
  \caption{Recoil mass against the $\ep\el$ pair for the data 
    sample after preselection cuts. The first peak on the left corresponds to the $\eta$ mass. The second 
peak at $\sim 590$ MeV is due to $\ks\to\pi^+\pi^-$ events with a 
wrong mass assignment.}
  \label{Fig:Mmiss}
\end{figure}

\begin{figure}[!t]
  \begin{center}
   \begin{tabular}{cc}
      \hspace{-0.7cm}
    \epsfig{file=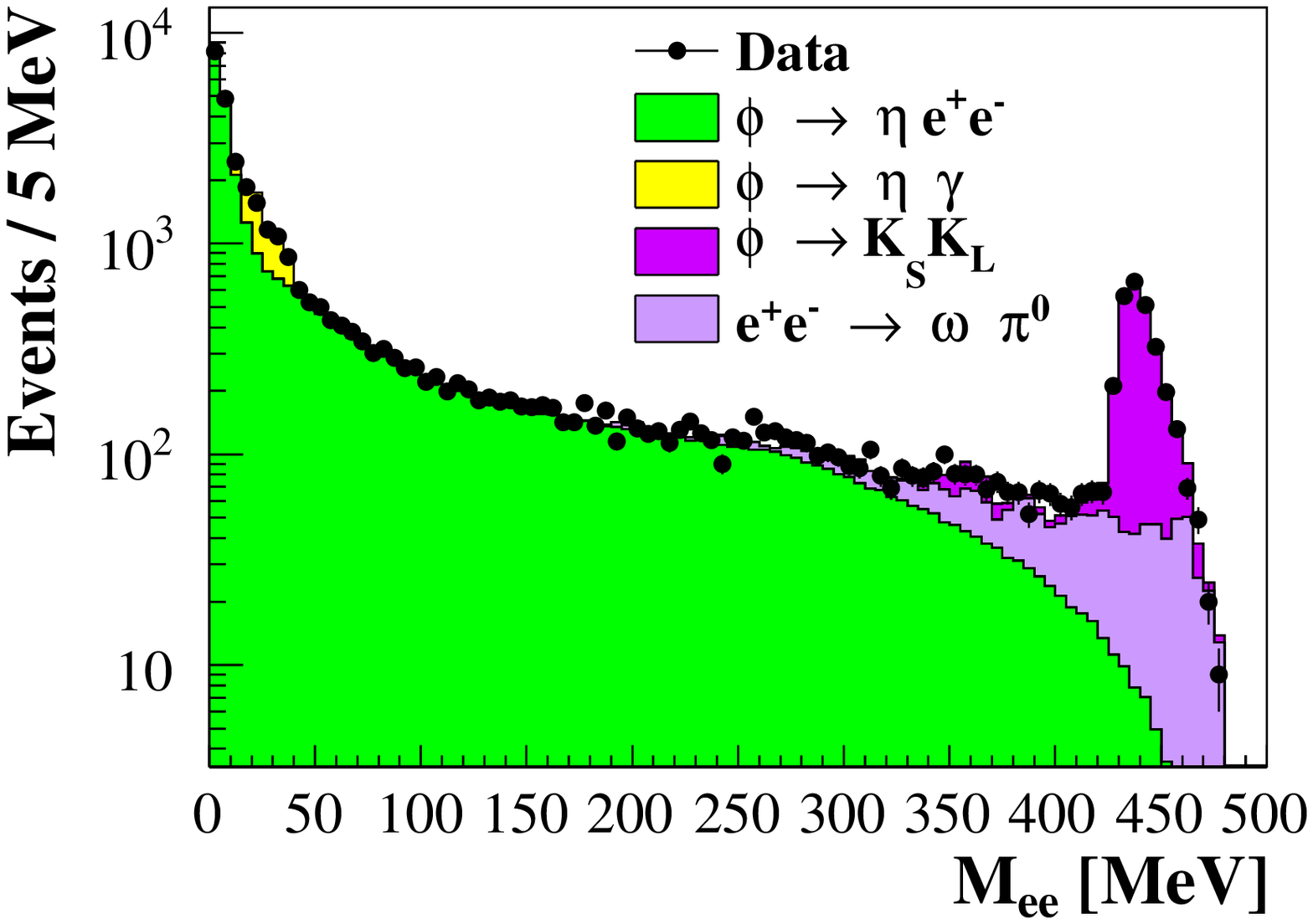,width=0.55\textwidth}
      \hspace{-0.9cm}
    \epsfig{file=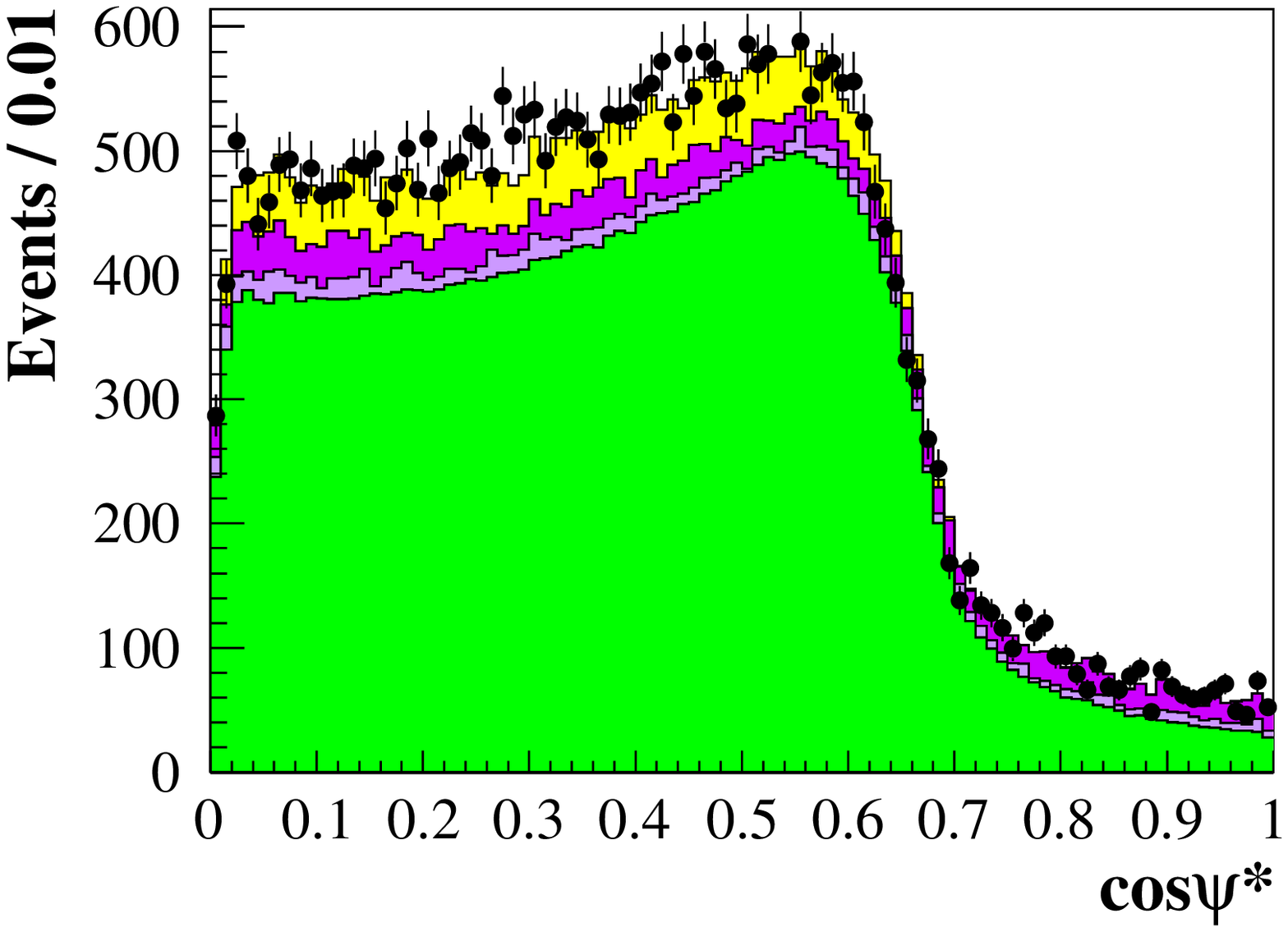,width=0.55\textwidth}\\
      \hspace{-0.7cm}
    \epsfig{file=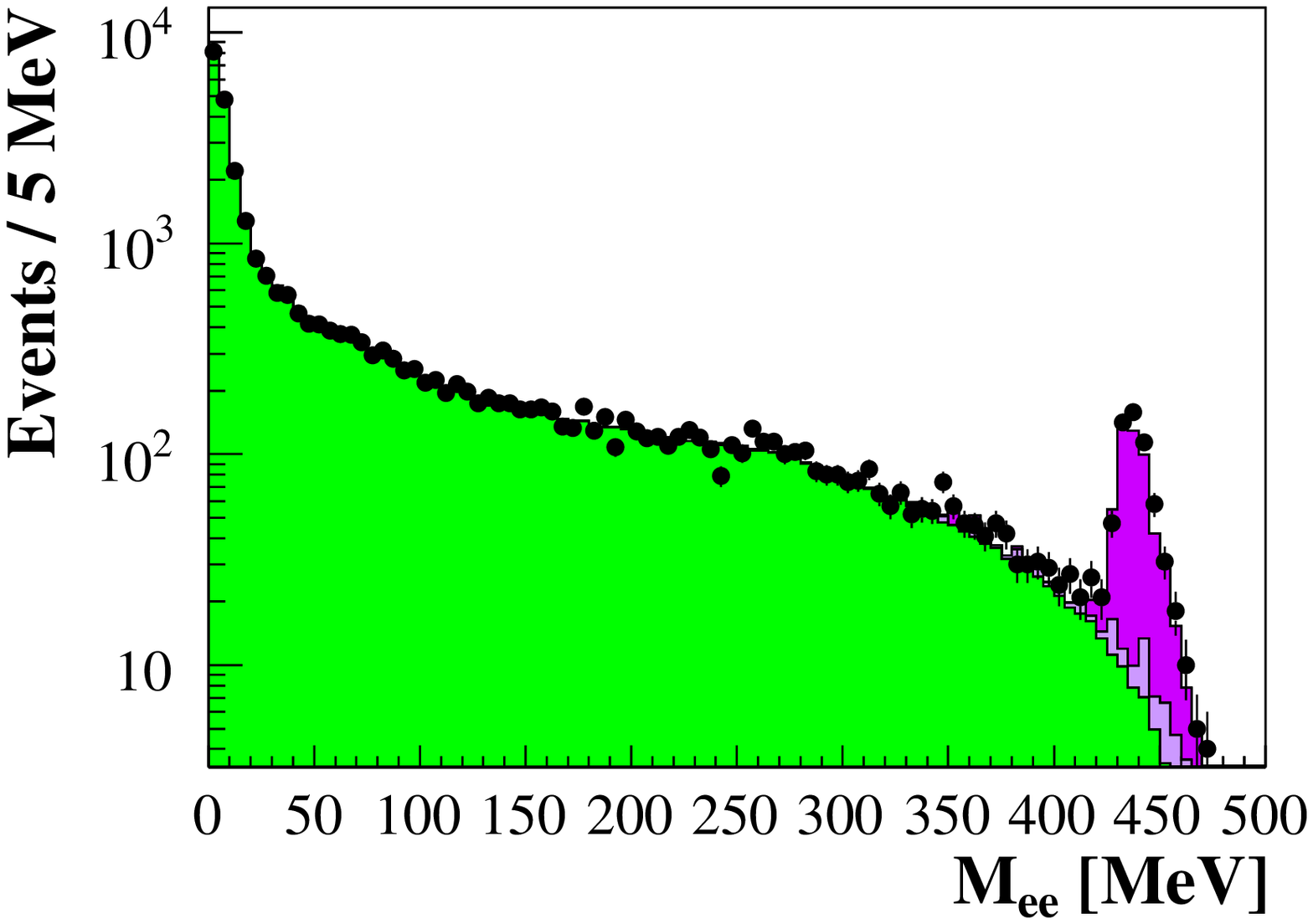,width=0.55\textwidth}
       \hspace{-0.9cm}
    \epsfig{file=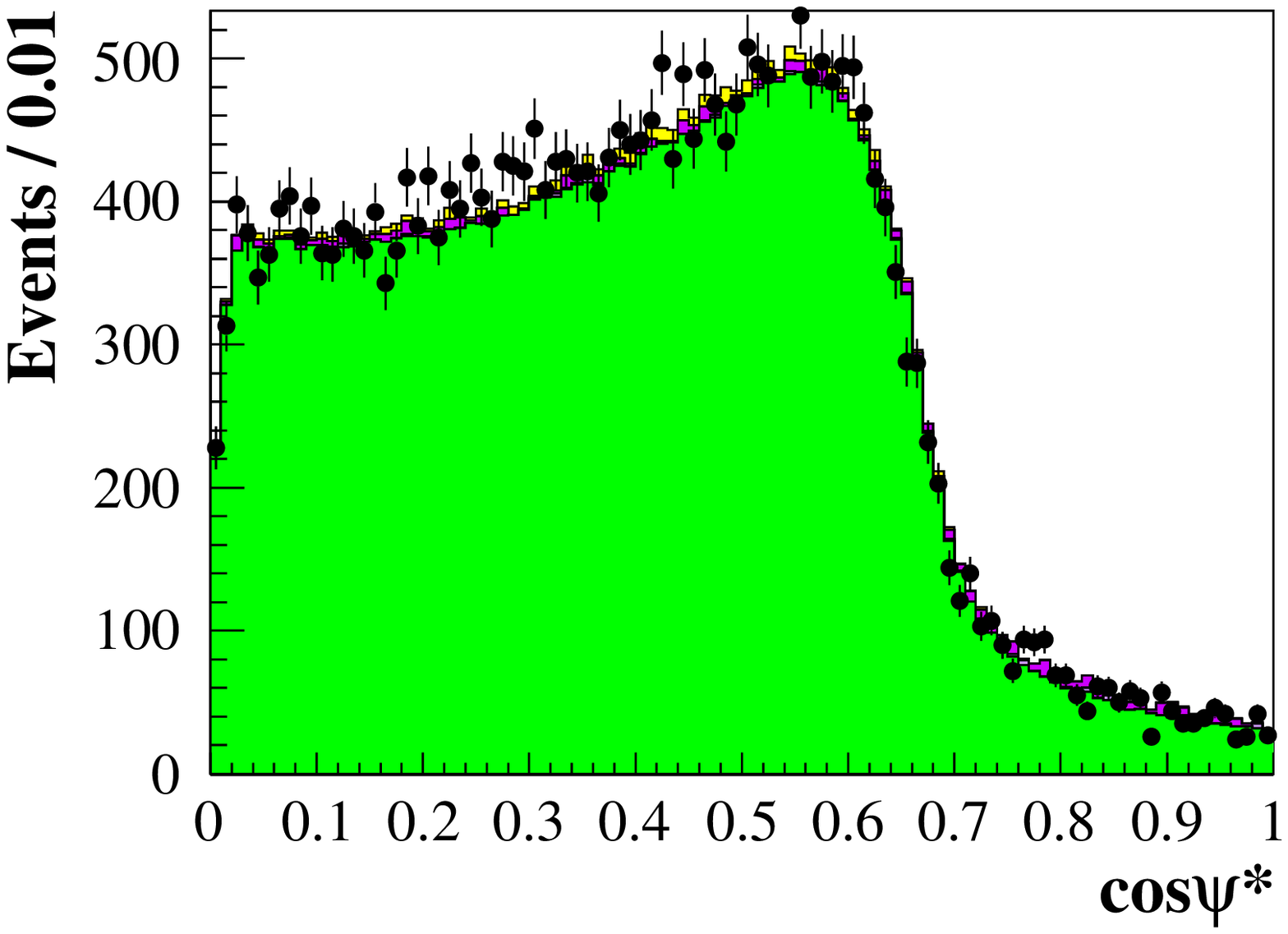,width=0.55\textwidth}    
     \end{tabular}
  \end{center}
  \caption{Data-MC comparison for $M_{ee}$ (left) and $\cos\psi^*$ (right)
    distributions after the $\rm{M_{ee}}(recoil)$ cut (top) and at the end of the analysis chain (bottom). The signal production corresponds to an integrated luminosity one hundred times larger than collected data.}
  \label{Fig:CompNeuBefore}
\end{figure}

The analysis of the decay chain $\phi \to \eta e^+e^-$, $\eta \to 3\pi^0$, has been performed on a data sample of about 1.7 fb$^{-1}$. The Monte Carlo (MC) simulation for the signal has been produced with $d\Gamma(\phi \to \eta e^+ e^-)/dM_{ee}$ according to VMD model. 
%and has been inserted in the KLOE simulation program, GEANFI \cite{biblio:geant_phi}. 
The signal production corresponds to an integrated luminosity one hundred times larger than collected data. Final state radiation has been included using PHOTOS Monte Carlo generator~\cite{biblio:Photos}. For the background, all $\phi$ decays and the not resonant $e^+e^-\to \omega \pi^0$ process have been simulated with a statistics two times larger than data.\\ 
All MC productions take into account changes in \dafne\ operation and background conditions on a run-by-run basis. Data-MC corrections for cluster energies and tracking efficiencies are evaluated with radiative Bhabha and $\phi\to\rho\pi$ samples, respectively.
%The selection cuts are based on the analysis described in~\cite{biblio:Uboson} for the search of a light dark boson with the same data sample. Here we summarise the main steps:
%The preselection has been performed as described in \cite{biblio:Uboson}. 
%As first analysis step for the neutral $\eta$ decay channel, a 
%preselection is performed requiring:
The main steps of the analysis are:
\begin{enumerate}
\item a preselection requiring two tracks of opposite sign extrapolated to a cylinder around the interaction point and 6 prompt photon candidates; 
\item a loose cut on the six photon invariant mass:  $400 < \rm{M_{6\gamma}} < 700$ MeV;
\item a 3$\sigma$ cut on the recoil mass against the $e^+e^-$ pair, $\rm{M_{ee}}(recoil)$, shown in Fig.~\ref{Fig:Mmiss}: $536.5 < \rm{M_{ee}}(recoil) < 554.5$ MeV\footnote{We observed a shift of about 2 MeV with respect to the $\eta$ mass ($\sim$ 547.85 MeV). The shift is due to the treatment of the energy loss for the electrons in the tracking reconstruction, that assumes the energy loss for pions.};
\item a cut on the invariant mass and the distance between the two tracks extrapolated to the beam pipe and at the drift chamber wall surfaces, to reject photon conversion;
\item a cut based on the time of flight (TOF) of the tracks to the calorimeter to reject events with charged pions in the final state.
\end{enumerate} 
%In Fig.~\ref{Fig:CompNeuBefore} (top), the comparison between data and Monte Carlo events for $M_{ee}$ and $\cos\psi^*$ distributions after the $M_{\rm recoil}(ee)$ cut is shown.\\
%The residual background contamination is mainly due to:
%\begin{enumerate}
%\item the photon conversions on the beam pipe (BP) or on the drift chamber walls (DCW), simulating an $e^+e^-$ pair from the interaction point;
%\item events with two charged pions in the final state. These are suppressed using the Time of Flight of tracks to the calorimeter.
%\end{enumerate}
These cuts are described in details in ref.~\cite{biblio:Uboson}, which reports the results for a search of a light vector boson using the same data sample. The $M_{ee}$ and $\cos\psi^*$\footnote{The $\cos\psi^*$ variable is defined as the angle between the $\eta$ and the $e^+$ in the \ep\el\ rest frame.} distributions, after the $\rm{M_{ee}}(recoil)$ cut and at the end of the analysis chain, are shown in Fig.~\ref{Fig:CompNeuBefore}, compared to MC expectations. The residual background contamination is concentrated at high masses and is dominated by $\phi \to K_SK_L\to \pi^+\pi^-3\pi^0$ events with an early $K_L$ decay.

\begin{figure}[!t]
  \begin{center}
   \epsfig{file=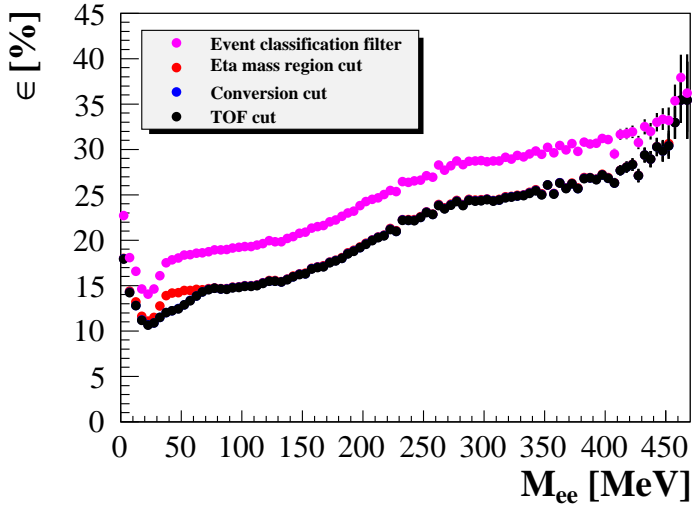, width=0.8\textwidth}
  \end{center}
  \caption{Analysis efficiency as a function
    of \ep\el\ invariant mass for different steps of the selection procedure.}
  \label{fig:effi}
\end{figure}

The analysis efficiency for signal events as a function of the $e^+e^-$ invariant mass is shown in Fig. \ref{fig:effi} for 5 MeV mass bins. It is about
%, defined as the ratio between the number of
%events surviving analysis cuts and that of all generated events, 
10$\%$ at low masses and increases to $\sim 35\%$ 
at 460 MeV, due to the larger acceptance for higher momentum tracks. 

At the end of the analysis chain, 30,577 events are selected, with 
$\sim 3\%$ background contamination.
After  bin to bin background subtraction, 29,625$\pm$178 $\phi\to\eta e^+e^-$, $\eta \to 3\pi^0$, 
candidates are present in the dataset.

%\begin{figure}[!t]
%  \begin{center}
%    \epsfig{file=fig_new/comp_after_mee.eps,width=0.8\textwidth}
%    \epsfig{file=fig_new/comp_after_cospsi.eps,width=0.8\textwidth}
%  \end{center}
%  \caption{$\phi\to\eta\,\ep\el$, $\eta\to\pio\pio\pio$ events: 
%    data-MC comparison for $M_{ee}$ (top) and $\cos\psi^*$ 
%    distributions (bottom) at the end of the analysis chain.}
%  \label{Fig:CompNeuAfter}
%\end{figure}

The branching ratio has been calculated using bin-by-bin efficiency  correction:
\begin{equation}
BR(\phi\to\eta e^+e^-) = \frac{\sum_i N_i/\epsilon _i}{\sigma_{\phi}\times \mathcal{L} \times BR(\eta \to 3 \pi^0)}.
\label{eq:BR}
\end{equation}
The luminosity measurement is obtained using very large angle Bhabha scattering events~\cite{biblio:boh_lumi_kloe}, giving an integrated luminosity of $\rm{\mathcal{L}=(1.68\pm0.01)\;fb^{-1}}$. The effective $\phi$ production cross section takes into account the center of mass energy variations (at 1\% level)~\cite{biblio:boh_cross_sec_kloe}: $\sigma\rm{=(3310\pm120)\;nb}$. The value of the BR($\eta \to 3\pi^0$)=(32.57$\pm$0.23)\% is taken from~\cite{biblio:pdg}. Our result is: 
\begin{equation}
% BR(\phi\to\eta e^+e^-)=(1.075\pm0.007\pm0.038^{+0.006}_{-0.002})\times 10^{-4},
 BR(\phi\to\eta e^+e^-)=(1.075\pm0.007\pm0.038)\times 10^{-4},
 \label{eq:BR2}
\end{equation}
where the error includes the uncertainties on luminosity and $\phi$ production cross section. The systematic error has been evaluated moving by  $\pm 1 \sigma$ the analysis cuts on the recoil mass and TOF, and by $\pm$ 20\% those related to conversion cuts (Table \ref{tab:syst_BR}). In order to evaluate the systematic due to the
variation of the analysis efficiency for low  $M_{ee}$ values, the BR has been measured for M$_{ee}> 100$ MeV, where the efficiency has a smoother behaviour. These systematics are negligible with respect to the normalization error.
\begin{table}
\caption{Systematics on the branching ratio. Relative variation of each contribution with respect to the $\rm{M_{ee}}(recoil)$, TOF, Photon Conversion, Event Classification cuts are reported.}
\centering
\small
\begin{tabular}{c|c|c}
  \textbf{CUT}  &        & \textbf{BR Variation} \\  
  \hline
%  \textbf{$M_{REC.}+1\sigma$} & -0.1\% \\
%  \textbf{$\;\;\;\;\;\;\;\;\;\;-1\sigma$} & +0.6\% \\
%  \hline
%  \textbf{$TOF+1\sigma$} & +0.01\% \\
%  \textbf{$\;\;\;\;\;\;\;\;\;\;-1\sigma$} & -0.1\% \\  
%  \hline
%  \textbf{$Conv. (small\;box)$} & -0.1\% \\
%  \textbf{$\;\;\;\;\;\;\;\;\;\;(large\;box)$} & +0.1\% \\  
%  \hline
%  \textbf{$ECL$} & -0.1\% \\  
 $\rm{M_{ee}}(recoil)$ & \textbf{$\pm1\sigma$} & (-0.1/+0.06)\% \\
  \hline
 TOF & \textbf{$\pm1\sigma$} & (+0.01/-0.1)\% \\
  \hline
 Photon conversion &   \textbf{$\pm20\%$}  & (-0.1/+0.1)\% \\
  \hline
 Event Classification &   \textbf{$\rm{M_{ee}>100\, MeV}$}  & -0.1\% \\  
  \hline
  & \textbf{Total} & \textbf{(-0.2/0.6)\%}  \\
\end{tabular}
\label{tab:syst_BR}
\end{table}

%==============================================================================
\section{\boldmath Measurement of the electromagnetic transition form factor}
\begin{figure}[!h]
   \begin{center}
     \epsfig{file=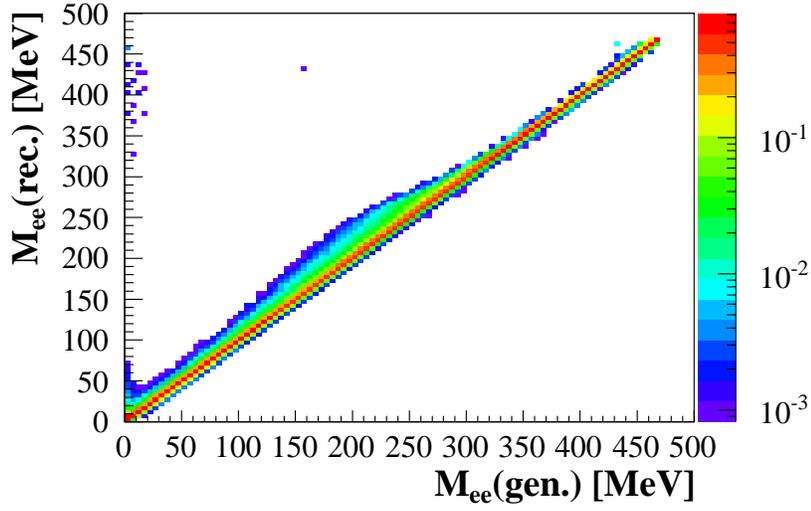,width=0.8\textwidth}
      \end{center}
     \caption{Smearing matrix:  reconstructed vs generated $M_{ee}$ values for $\phi \to \eta e^+e^-$ MC events.}
     \label{fig:smearing}
 \end{figure}

%==============================================================================
The fit procedure, based on the MINUIT package~\cite{biblio:minuit4}, is applied to the $M_{ee}$ distribution, after a bin-by-bin background subtraction. Analysis efficiency and smearing effects have been folded into the theoretical function of Eq.~\ref{eq:fit_func}, 
using as free parameters $\Lambda_{\phi \eta}$ with an overall normalization factor. 
The $M_{ee}$ distribution is then fitted, in the whole range, using a bin width of 5 MeV, by minimizing a $\chi^2$ function, defined as:
\begin{equation}
\chi^2=\sum_{i=1}^N \frac{(N_{DATA}^i - N_{expected}^i)^2}{\sigma _i^2},
\end{equation}  

where $\rm{N_{DATA}}$ is the number of event in the reconstructed i-th $M_{ee}$ bin after background subtraction and $\rm{N_{expected}}$ is the expected number of events in the same bin, evaluated by performing a convolution of the theoretical function with reconstruction effects as follows:
\begin{equation}
\label{eq:teoretical_func}
N_{expected}^i=\sum_{j=1}^{N} f_{theor.}(m_j) \cdot p(M_{ee}^j,M_{ee}^i)\cdot \epsilon_j,
\end{equation}  
where f$\rm{_{theor.}}$(m$_j$) is the integrated VMD spectrum in the j-th bin, p(m$_{ee}^j$,m$_{ee}^i$) is the probability for an events generated with mass m$_j$ to be reconstructed in the i-th bin and $\epsilon_j$ is the reconstruction efficiency in the j-th bin. The probability p(m$_{ee}^j$,m$_{ee}^i$) is shown in Fig.~\ref{fig:smearing}. 
%using 1 MeV binning in j, except for the first reconstructed $M_{ee}$ bin, when $\rm{\Delta m_j=0.5}$ MeV has been used. In Eq.~\ref{eq:teoretical_func} $\rm{dN/d(m_j)}$ is the $M_{ee}$ spectrum, $\rm{p(M_{ee}^j,M_{ee}^i)}$ is the probability for an event with mass $\rm{M_{ee}^j}$ to be reconstructed in i-th bin, as shown in Fig.~\ref{fig:smearing}. 
Smearing effects are of the order of few \%. The resolution on the $M_{ee}$ variable has been evaluated for each mass bin applying a gaussian fit on the $\rm{M_{ee}(rec.)-M_{ee}(true)}$ and it is at the 2\% level. 

\begin{figure}[!t]
  \begin{center}
    \epsfig{file=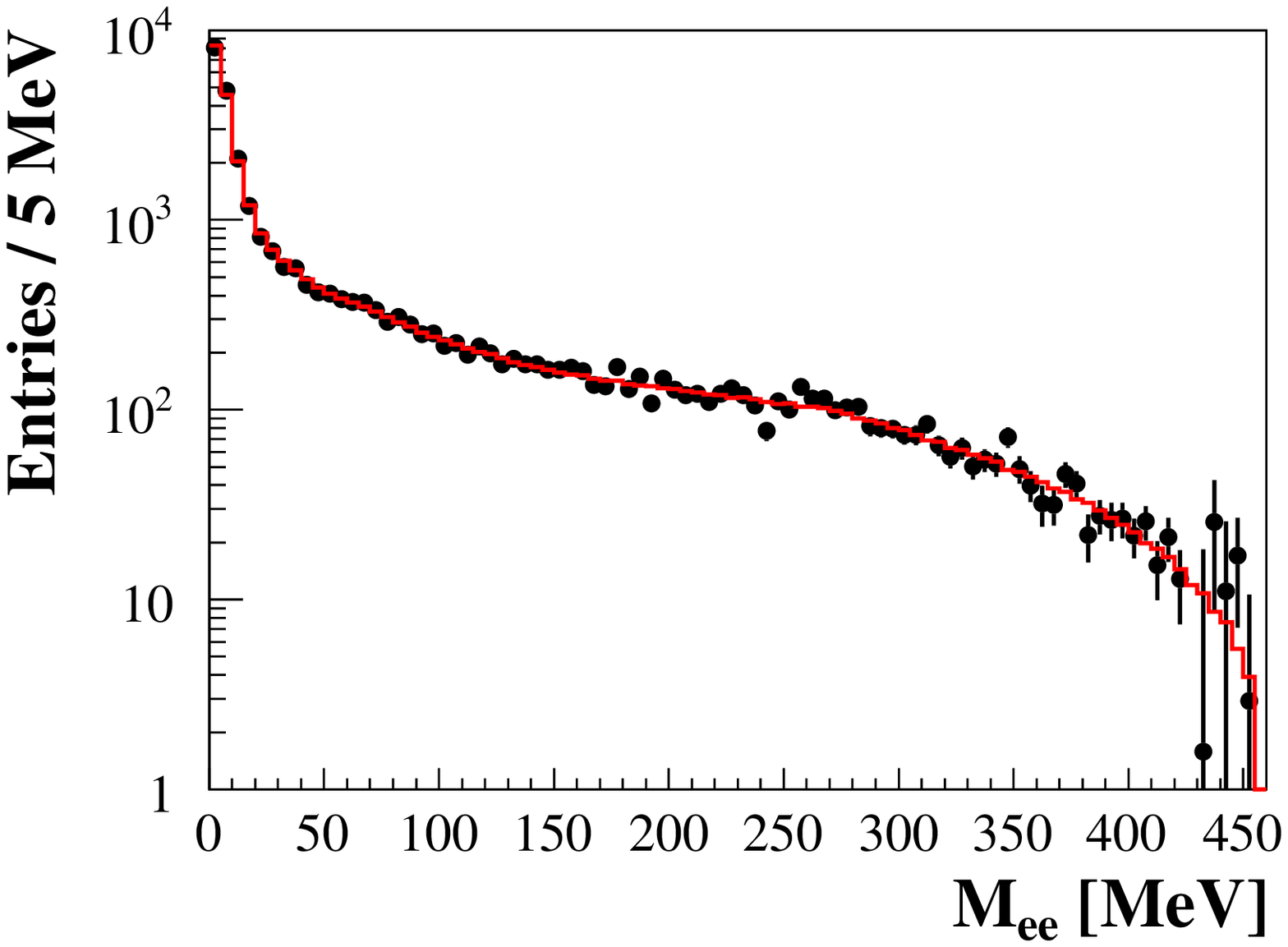,width=0.8\textwidth}
    \epsfig{file=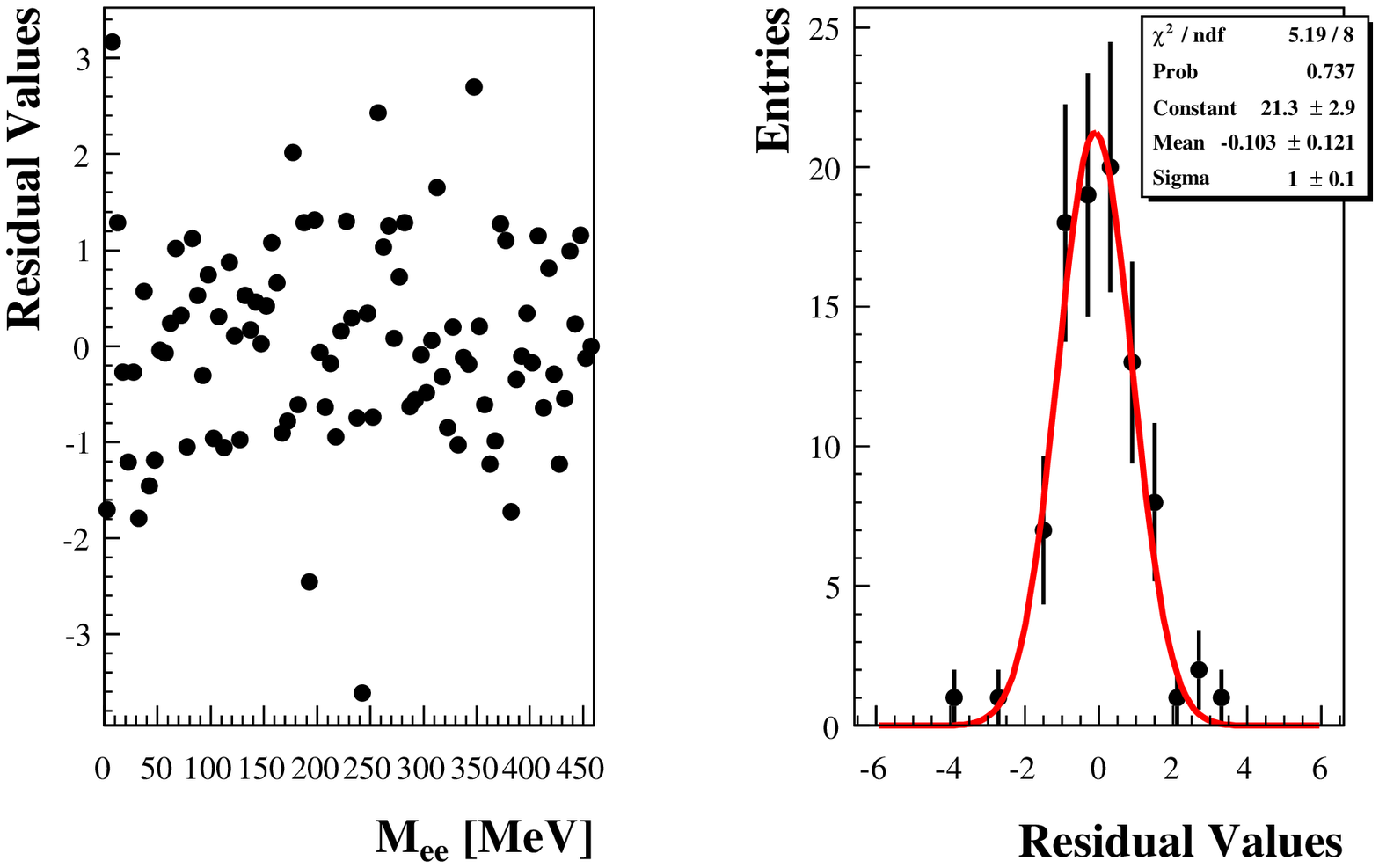,width=0.8\textwidth}
  \end{center}
  \caption{Top: fit to the $M_{ee}$ spectrum for the Dalitz 
    decays $\phi\to\eta\,e^+e^-$, with $\eta\to\pio\pio\pio$, in logarithmic scale. Bottom left: normalized fit residuals  vs $M_{ee}$. Bottom right: distribution of normalized values with superimposed a gaussian fit.}
  \label{Fig:Fit}
\end{figure}

As result of the fit procedure, we determine a value of the form factor slope $\rm{b_{\phi \eta}=(1.17\pm0.10)\;\;GeV^{-2}}$,
with $\chi^2/{\rm ndf}=1.17$ and a $\chi^2$ probability of about 13\%. 
%The value found for the normalization parameter is (108.6$\pm$0.7)$\times$10$^6$ with a correlation of 39\%. 
In Fig.~\ref{Fig:Fit} (top) the fit result is shown and compared with data. Fit normalized residuals, defined as $\rm{(N_{DATA}^i - N_{expected}^i)/\sigma _i}$,  are shown in Fig.~\ref{Fig:Fit} bottom left: the distribution of their values has the correct gaussian behaviour, centered at 0 with $\sigma =1$ (Fig.~\ref{Fig:Fit} bottom right). \\
%This result is in good agreement with the VDM prediction, and improves the SND measurement \cite{biblio:snd4} by a factor of about five.\\
Systematics for the $\rm{M_{ee}}(recoil)$, TOF and photon conversion cuts have been evaluated as for the BR measurement and summarised in Table~\ref{tab:all_sys}. Systematics related to the fit procedure have been evaluated as the RMS of the deviation from the central value obtained by varying the mass range used for the fit. The total systematic error is the quadrature of all contributions. \\
The result for the slope of the transition form factor is:
\begin{equation}
\label{eq:FIT1}
\rm{b_{\phi \eta}=(1.17\pm0.10^{+0.07}_{-0.11})\;\;GeV^{-2}}.
\end{equation}

 \begin{table}[!h]
\centering
\caption{Systematics on $b_{\phi \eta}$. Relative variation of each contribution with respect to the $\rm{M_{ee}}(recoil)$, TOF, Photon Conversion, Fit mass range cuts are reported.}
\small
\begin{tabular}{c|c|c}
  \textbf{CUT} & & \textbf{b$_{\phi\eta}$ Variation} \\  
  \hline
 $\rm{M_{ee}}(recoil)$ & \textbf{$\pm1\sigma$} & (+3.3/-4.6)\% \\
  \hline
 TOF & \textbf{$\pm1\sigma$} & (-2.5/1.5)\% \\
  \hline
 Photon conversion &   \textbf{$\pm20\%$}  & (-5.9/1.7)\% \\
  \hline
 Fit Limits &  $M_{ee}$ fit range   & $\pm$4.4\% \\  
  \hline
 &  \textbf{Total}  & \textbf{(-9.0/+6.0)\%}  \\
\end{tabular}
\label{tab:all_sys}
\end{table}

%
%==============================================================================
 \section{Transition form factor as a function of M$\mathbf{_{ee}}$}
%==============================================================================
The modulus squared of the transition form factor, $|F_{\phi\eta}(q^2)|^2$, as a function of the $e^+e^-$ invariant mass, is obtained by dividing bin by bin the $M_{ee}$ spectrum of Fig.~\ref{Fig:Fit} (top) by the one of reconstructed signal events, generated with $F_{\phi \eta}^{MC}=1$, after all analysis cuts.
%distribution (5 MeV binning) obtained by the MC simulated with $\rm{F_{\phi \eta}^{MC}=1}$, after all analysis cut.
MC sample is normalized in order to reproduce the number of events in the first bin of data. In Table~\ref{tab:mee_point}, the values of $|F_{\phi\eta}(q^2)|^2$ as a function of the dilepton invariant mass, with the corresponding statistical errors are reported. 
\begin{table}[h]
\caption{Transition form factor $|$F$_{\phi \eta}$$|$$^2$ of the $\phi \to \eta e^+e^-$ decay.} 
\tiny
\begin{tabular}{lclclclclclclclc|c|c|c|}
$M_{ee}$ (MeV) & $|$F$_{\phi \eta}$$|$$^2$ & $\delta|$F$_{\phi \eta}$$|$$^2$ & \;\;\;\;\;\; & $M_{ee}$ (MeV)  & $|$F$_{\phi \eta}$$|$$^2$ & $\delta|$F$_{\phi \eta}$$|$$^2$ & \;\;\;\;\;\; & $M_{ee}$ (MeV)  & $|$F$_{\phi \eta}$$|$$^2$ & $\delta|$F$_{\phi \eta}$$|$$^2$\\ 
\hline
\hline
2.50 & 1.00 & 0.01    &  &     157.50 & 1.17 & 0.09  &  &  312.50 & 1.57 & 0.17 \\
7.50 & 1.05 & 0.02    &  &     162.50 & 1.13 & 0.09  &  &  317.50 & 1.28 & 0.16 \\		  
12.50 & 1.03 & 0.02   &  &     167.50 & 0.98 & 0.08  &  &  322.50 & 1.19 & 0.16 \\		  
17.50 & 0.99 & 0.03   &  &     172.50 & 1.03 & 0.09  &  &  327.50 & 1.38 & 0.18 \\		   
22.50 & 0.97 & 0.04   &  &     177.50 & 1.28 & 0.10  &  &  332.50 & 1.21 & 0.18 \\		  
27.50 & 1.00 & 0.04   &  &     182.50 & 1.03 & 0.09  &  &  337.50 & 1.35 & 0.19 \\		  
32.50 & 0.93 & 0.04   &  &     187.50 & 1.21 & 0.10  &  &  342.50 & 1.39 & 0.20 \\		  
37.50 & 1.03 & 0.05   &  &     192.50 & 0.90 & 0.09  &  &  347.50 & 2.08 & 0.26 \\		  
42.50 & 0.95 & 0.05   &  &     197.50 & 1.25 & 0.10  &  &  352.50 & 1.50 & 0.25 \\		  
47.50 & 0.95 & 0.05   &  &     202.50 & 1.12 & 0.10  &  &  357.50 & 1.30 & 0.24 \\		  
52.50 & 1.01 & 0.05   &  &     207.50 & 1.05 & 0.10  &  &  362.50 & 1.13 & 0.28 \\		  
57.50 & 1.01 & 0.05   &  &     212.50 & 1.13 & 0.10  &  &  367.50 & 1.20 & 0.27 \\		  
62.50 & 1.03 & 0.05   &  &     217.50 & 1.04 & 0.10  &  &  372.50 & 1.87 & 0.29 \\		  
67.50 & 1.08 & 0.06   &  &     222.50 & 1.14 & 0.10  &  &  377.50 & 1.76 & 0.29 \\		  
72.50 & 1.04 & 0.06   &  &     227.50 & 1.27 & 0.11  &  &  382.50 & 1.02 & 0.29 \\
77.50 & 0.96 & 0.06   &  &     232.50 & 1.18 & 0.11  &  &  387.50 & 1.49 & 0.31 \\
82.50 & 1.09 & 0.06   &  &     237.50 & 1.06 & 0.10  &  &  392.50 & 1.58 & 0.36 \\						
87.50 & 1.06 & 0.06   &  &     242.50 & 0.83 & 0.10  &  &  397.50 & 1.79 & 0.38 \\
92.50 & 1.01 & 0.06   &  &     247.50 & 1.20 & 0.11  &  &  402.50 & 1.54 & 0.37 \\
97.50 & 1.08 & 0.07   &  &     252.50 & 1.11 & 0.11  &  &  407.50 & 2.08 & 0.43 \\
102.50 & 0.98 & 0.07  &  &     257.50 & 1.52 & 0.13  &  &  412.50 & 1.40 & 0.48 \\
107.50 & 1.06 & 0.07  &  &     262.50 & 1.33 & 0.12  &  &  417.50 & 2.24 & 0.59 \\
112.50 & 0.97 & 0.07  &  &     267.50 & 1.39 & 0.13  &  &  422.50 & 1.40 & 0.59 \\
117.50 & 1.12 & 0.08  &  &     272.50 & 1.24 & 0.13  &  &  427.50 & -0.14& 1.36 \\
122.50 & 1.05 & 0.08  &  &     277.50 & 1.32 & 0.13  &  &  432.50 & 0.28 & 3.02 \\
127.50 & 0.96 & 0.07  &  &     282.50 & 1.39 & 0.14  &  &  437.50 & 5.36 & 3.59 \\
132.50 & 1.09 & 0.08  &  &     287.50 & 1.18 & 0.13  &  &  442.50 & 2.75 & 3.68 \\
137.50 & 1.06 & 0.08  &  &     292.50 & 1.20 & 0.13  &  &  447.50 & 6.97 & 4.10 \\
142.50 & 1.08 & 0.08  &  &     297.50 & 1.27 & 0.14  &  &  452.50 & 1.44 & 3.79 \\
147.50 & 1.06 & 0.08  &  &     302.50 & 1.22 & 0.14  &  &  457.50 & 3.43 & 4.91 \\
152.50 & 1.11 & 0.09  &  &     307.50 & 1.30 & 0.15  &  &  	            &	      &         \\
\end{tabular}
\label{tab:mee_point}
\end{table}
\normalsize

\begin{figure}[!h]
   \begin{center}
     \epsfig{file=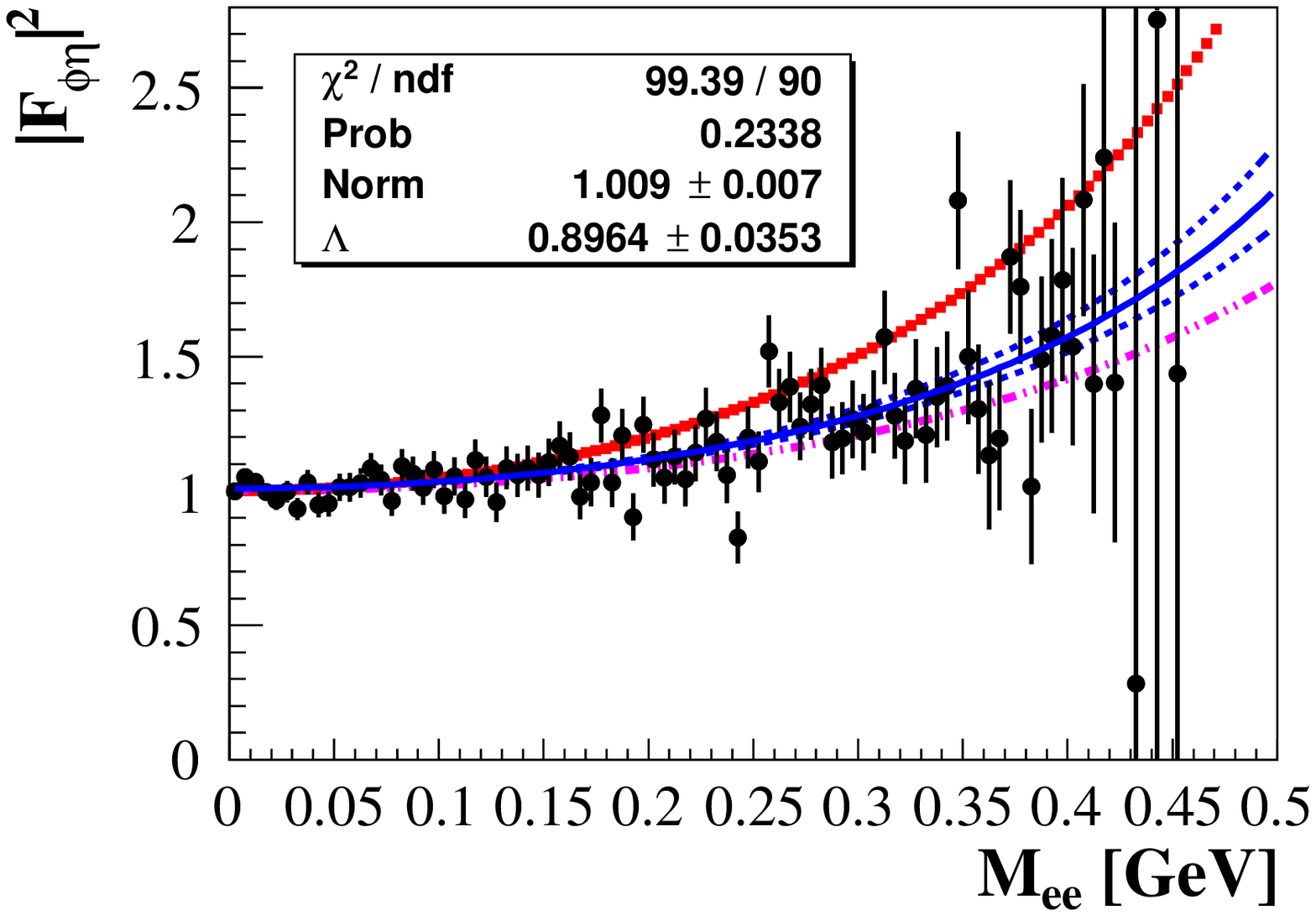,width=0.8\textwidth}
     \caption{Fit to the $|$F$_{\phi \eta}$$|$$^2$ distribution as a function of the invariant mass of the electron positron pair, with a binning of 5 MeV. The blue curve is the fit result, and in dashed blue the functions obtained for $\Lambda_{\phi\eta}$=$\Lambda_{\phi\eta} \pm 1\sigma$ are reported. VMD expectations are superimposed in pink while the curve obtained from reference~\cite{biblio:Leupold} is reported in red.} 
     \label{fig:data_5MeV}
   \end{center}
 \end{figure}

The $|F_{\phi\eta}(q^2)|^2$ distribution has been fitted as a function of the invariant mass with two free parameters, one corresponding to the normalization and the other to $\Lambda_{\phi\eta}$, as shown in Fig.~\ref{fig:data_5MeV}, together with the predictions form the VMD and from ref.~\cite{biblio:Leupold}. From this fit, the value of the slope $b_{\phi \eta}$ is:
\begin{equation}
b_{\phi \eta}=(1.25\pm0.10)\;\;\rm{GeV^{-2}},
\label{eq:FIT_MASS}
\end{equation}
\noindent in agreement within the uncertainties with the value obtained from the fit to the invariant mass spectrum (Eq.~\ref{eq:FIT1}).

 %\clearpage
%==============================================================================
\section{\boldmath Conclusions}
%==============================================================================
Analysing  the $\phi \to \eta e^+e^-$ decay channel, an precise measurements of both, the BR($\phi \to \eta e^+e^-$), and the transition form factor slope $b_{\phi\eta}$ are obtained. We measured a value of $\rm{BR(\phi \to \eta \mathit{e^+e^-}}) = $$\rm{(1.075 \pm 0.007 \pm 0.038)}$$\rm{ \times 10^{-4}}$ and a value of the slope of $\rm{b_{\phi \eta} = }$$\rm{(1.17 \pm 0.10 ^{+0.07}_{-0.11})\,GeV^{-2}}$. \\
The BR($\phi\to\eta e^+e^-$) is in agreement with VMD predictions~\cite{biblio:vmd_expectation} and with the SND and CMD-2 results~\cite{biblio:snd4, biblio:cmd4}. The transition form factor slope is in agreement with VMD predictions~\cite{biblio:vmd_expectation}, with a precision that is a factor of five better than previous SND measurement.\\
%\begin{table}[h!]
%\caption{Decay probability and form factor slope results compared
%with VDM predictions and previous experiments.}
%\scriptsize
%\begin{tabular}{c|c|c|c}
%                                                              &  \textbf{This work}                                                        &  \textbf{VMD prediction} \cite{biblio:vmd_expectation}  &  \textbf{Other experiments}\\
%\hline
%$\rm{BR(\times 10^4)}$                         &  $\rm{1.075\pm0.007\pm0.038^{+0.006}_{-0.002}}$    &  1.1                                                                               &  
%$\rm{1.19 \pm 0.19 \pm 0.07}$ \cite{biblio:snd4}\\ 
% & & & $\rm{1.14 \pm 0.10 \pm 0.06}$ \cite{biblio:cmd4}\\
%
%\hline
%$\rm{b_{\phi \eta}}$($\rm{GeV^{-2}}$)   &  $\rm{1.17\pm0.10^{+0.07}_{-0.11}}$                           & 1.0										    &
%$\rm{3.8\pm 1.8}$ \cite{biblio:snd4}\\
%
%\hline  
%\end{tabular}
%\label{Tab:Result}
%\end{table}%
The transition form factor has been used~\cite{biblio:Wang} to derive the upper limit for the production of a light dark boson U in $\phi \to \eta U \to \eta e^+e^-$ decay. Present measurement confirms the exclusion plot obtained by KLOE in the mass range (5 $<$ M$_U$ $<$ 470) MeV, where b$_{\phi \eta}$ = 1 GeV$^{-2}$ was assumed~\cite{biblio:Uboson}.

%======================================================================
\section*{Acknowledgments}
%======================================================================

We warmly thank our former KLOE colleagues for the access to the data collected during the KLOE data taking campaign.
We thank the DA$\Phi$NE team for their efforts in maintaining low background running conditions and their collaboration during all data taking. We want to thank our technical staff: 
G.F. Fortugno and F. Sborzacchi for their dedication in ensuring efficient operation of the KLOE computing facilities; 
M. Anelli for his continuous attention to the gas system and detector safety; 
A. Balla, M. Gatta, G. Corradi and G. Papalino for electronics maintenance; 
M. Santoni, G. Paoluzzi and R. Rosellini for general detector support; 
C. Piscitelli for his help during major maintenance periods. 
This work was supported in part by the EU Integrated Infrastructure Initiative Hadron Physics Project under contract number RII3-CT- 2004-506078; by the European Commission under the 7th Framework Programme through the `Research Infrastructures' action of the `Capacities' Programme, Call: FP7-INFRASTRUCTURES-2008-1, Grant Agreement No. 227431; by the Polish National Science Centre through the Grants No. 
%0469/B/H03/2009/37, 
%0309/B/H03/2011/40, 
DEC-2011/03/N/ST2/02641, 
2011/01/D/ST2/00748,
2011/03/N/ST2/02652,\\
2013/08/M/ST2/00323,
and by the Foundation for Polish Science through the MPD programme and the project HOMING PLUS BIS/2011-4/3.

%======================================================================

%%%%%%%%%%%%%%%%%%%%%%%%%%%%%%%%%%%%%%%%%%%%%%%%%%%%%%%%%%%%%%%%%%%%%%%%%%%%%%%
\end{document}